\documentclass[12pt,preprint]{aastex}

\slugcomment{Accepted by ApJL. Dec. 5th, 2005}

\begin{document}

\title{CO J=6--5 observations of TW~Hya with the SMA}

\author{Chunhua~Qi\altaffilmark{1}, David~J.~Wilner\altaffilmark{1},
Nuria~Calvet\altaffilmark{1,4}, Tyler~L.~Bourke\altaffilmark{1},
Geoffrey~A.~Blake\altaffilmark{2},
Michiel~R.~Hogerheijde\altaffilmark{3}, Paul~T.P.~Ho\altaffilmark{1,5},
Edwin~Bergin\altaffilmark{4}}
\altaffiltext{1}{Harvard--Smithsonian Center for
Astrophysics, 60 Garden Street, MS 42, Cambridge, MA 02138,
USA; cqi@cfa.harvard.edu, dwilner@cfa.harvard.edu,
ncalvet@cfa.harvard.edu, tbourke@cfa.harvard.edu, ho@cfa.harvard.edu}
\altaffiltext{2}{Divisions of Geological \& Planetary Sciences and
Chemistry \& Chemical Engineering, California Institute of Technology,
MS150--21, Pasadena, CA 91125, USA; gab@gps.caltech.edu.}
\altaffiltext{3}{Sterrewacht Leiden, P.O. Box 9513,
2300 RA Leiden, The Netherlands; michiel@strw.leidenuniv.nl.}
\altaffiltext{4}{Department of Astronomy, University of Michigan, 825
Dennison Building, 500 Church Street, Ann Arbor, MI 48109;
ncalvet@umich.edu, ebergin@umich.edu.}
\altaffiltext{5}{Academia Sinica Institute of Astronomy \& 
Astrophysics, P.O. Box 23-141, Taipei, Taiwan, 106, R.O.C.; ho@asiaa.sinica.edu.tw.}

\begin{abstract}
We present the first images of the 691.473 GHz CO J=6--5 line in a 
protoplanetary disk, obtained along with the 690 GHz dust continuum,
toward the classical T Tauri star TW~Hya using the Submillimeter Array.
Imaging in the CO J=6--5 line reveals a rotating disk, 
consistent with previous observations of CO J=3--2 and 2--1 lines. 
Using an irradiated accretion disk model and 2D Monte Carlo radiative 
transfer, we find that additional surface heating is needed to fit 
simultaneously the absolute and relative intensities of the CO J=6--5, 3--2 
and 2--1 lines. In particular, the vertical gas temperature gradient
in the disk must be steeper than that of the dust, mostly likely because
the CO emission lines probe nearer to the surface of the disk.
We have used an idealized  X-ray heating model to fit the line profiles 
of CO J=2--1 and 3--2 with $\chi^2$ analysis, and the prediction of this 
model yields CO J=6--5 emission consistent with the observations.
\end{abstract}


\section{Introduction}

Disks around pre-main sequence stars are the likely sites of the formation
of planetary systems (e.g. ~\citealp{beckwith99}). 
Our quantitative understanding of the physical properties of such disks --- 
in particular their sizes, radial and vertical temperature and density 
structures, and gas survival timescales --- has improved dramatically in 
recent years as (sub)millimeter-wave interferometers have imaged 
the gas and dust surrounding over a dozen T Tauri and Herbig Ae stars
at $\sim$1$''$--5$''$ resolution (eg. \citealp{koerner_s95,
guilloteau_d98, dartois_d03, qi_k03, qi_h04}).
Observations and modeling of molecular emission of trace species 
can be used to derive the density and temperature structure in disks. 
Although the structure inferred from observations of a single line 
is not unique, observations of a sufficiently large number of
lines of various molecules can be used to constrain the
temperature and density independently (van Zadelhoff et al. 2001). 

TW~Hya is the closest known classical T Tauri star (at a distance of
56 pc), and its circumstellar disk has been imaged in the CO J=3--2 
and J=2--1 lines with the SMA (Qi et al. 2004). A detailed comparison 
of the data with 2D Monte Carlo radiative transfer models was used 
to constrain
the disk properties. One problem with these models was that the 
predicted CO J=3--2 intensities were always lower than the observed values, 
no matter what disk parameters were varied (including outer disk radius, 
inclination angle and molecular depletion factors). 
Since the disk is externally irradiated, the gaseous component can be 
detectable through emission lines produced in temperature inversions 
of the disk atmosphere where CO lines are optically thick. 
The disk thermal structure is mainly derived from the fitting of the
spectral energy distribution (SED), which provides a good description 
of the dust temperature profile. However, the sampling of the disk temperature 
distribution by gas and dust is different, as the optically thin dust probes 
much deeper into the mid-plane and closer to the star in radius.
Qi et al. (2004) suggested that the difference between the CO intensity 
and the canonical model may be due to the higher gas temperature probed 
by the CO emission than the dust temperature. 
For this reason, we searched for CO J=6--5 emission, 
which is expected to be sensitive to the slope of the gas temperature
inversion at the surface of the disk.

We present here the first aperture synthesis mapping of CO
J=6--5 in a protoplanetary disk, using the Submillimeter Array (SMA).
We show that inclusion of additional disk surface gas heating in
the disk model, in the form of an idealized X-ray heating model, 
provides a better fit to both our previous SMA observations of the 
CO J=3--2 and 2--1 lines, and a much better fit to the observations
of the CO J=6--5 line, than the model without additional heating.

\section{Observations}

The 690 GHz observations of TW~Hya were made with the
SMA\footnote{The Submillimeter Array is a joint
project between the Smithsonian Astrophysical Observatory and the
Academia Sinica Institute of Astronomy and Astrophysics, and is
funded by the Smithsonian Institution and the Academia Sinica.}
(\citealp{ho_m04}) on 17 February 2005
in a compact configuration with six antennas under excellent sky 
conditions with $\tau_{225 Ghz}\sim0.03-0.05$. These
observations provided 15 independent baselines ranging in length from
16 to 156 meters. Table~\ref{tab:obs} summarizes the observational
parameters (additional CO J=2--1 data was obtained with longer
baselines than reported by Qi et al. 2004 and are also summarized here).
The synthesized beam sizes were $2\farcs4\times0\farcs9$ for 690 GHz
continuum emission and $3\farcs9\times1\farcs2$ for
CO J=6--5 with natural weighting. The SMA digital correlator was
configured with a narrow band of 128 channels
over 104 MHz, which provided 0.8 MHz frequency resolution, or
0.35 km~s$^{-1}$ velocity resolution at 690 GHz, and several broader
bands that together provided 2 GHz for continuum measurements.
Calibration of the visibility phases and amplitudes was achieved with
observations of Callisto, at intervals of typically 20 minutes. 
During the observations, Callisto was 40 degrees away from TW~Hya 
and had a diameter of $1\farcs39$ and a zero-spacing flux density of 
45.3 Jy at 690 GHz, which provided the absolute scale for the 
flux density calibration. The uncertainties in the flux scale are 
estimated to be 10\% according to the uncertainties of the Callisto model. 
The 4.6 Jy continuum emission from TW~Hya is strong enough for 
self-calibration, and one iteration of phase-only self-calibration
with the 690 GHz TW~Hya continuum model was performed on the CO J=6--5
data to improve the images. The MIRIAD package was used for imaging.

\section{Results}

Figure 1 shows the CO J=6--5 channel maps (one channel at velocity 2.9 km
s$^{-1}$ in red contours and a second channel at velocity 2.5 km s$^{-1}$
in blue contours) overlaid on the 690 GHz continuum image in gray scale. 
The velocity gradient along the disk position angle of $-$27 degrees 
is consistent with that seen in CO J=2--1 and J=3--2 (Qi et al. 2004).

The disk averaged CO J=6--5 spectrum, along with the CO J=3--2 and 2--1
spectra (in black lines), toward TW~Hya at $\sim$2$''$ resolution is 
presented in Figure 2. The spectra predicted by the canonical model 
from Qi et al. (2004) are shown in solid blue lines. The canonical model,
adapted from Qi et al. (2004) by fitting both CO J=3--2 and 2--1 data,
is summarized in Table 2 and discussed further below. 
As shown in Figure 2, the difference between the data and the model 
becomes substantial as the energy of the CO transitions increases.
This indicates a steeper vertical temperature profile in the gas than 
in the dust near the disk surface, which suggests an additional source 
of gas heating at the disk surface is required.

\citet{cg97} show that the thermal properties of the dust in the 
upper layers of a disk are uncoupled from those of the gas. 
The dust temperature is determined only by the absorption
of stellar radiation and the
re-radiation by the dust, while the gas temperature is
determined not only by its collisions with the dust but also by 
other gas heating and cooling processes. As indicated by 
Glassgold \& Najita (2001),
above a certain height, the gas and dust should be treated as
distinct (but interacting) thermodynamic systems, and existing disk
atmospheres do not give the the correct temperature for the gas.
As shown in Figure 2, the canonical model predictions are
weaker than the observations, and the CO 6--5 lines shows the 
largest discrepancy, which
is clear evidence that the canonical model needs to be adjusted.

\section{Discussion}

Both X-ray (\citealp{glassgold_n04}) and
ultraviolet (\citealp{Jonkheid_f04}) radiation are effective in heating
the gas in the surface layers of disks. 
Recent X-ray satellite observations confirm that essentially all young
stellar objects are luminous X-ray emitters (\citealp{feigelson_m99}).
Observational constraints on the various aspects of X-ray irradiation
of protoplanetary disks are rapidly improving, and it appears that the
emitted X-rays can affect the environment of a young star out to
hundreds of AU (\citealp{glassgold_n01}).
\citet{igea_g99} showed that Compton processes increase X-ray
penetration through column densities of order $10^{24}$-$10^{25}$
cm$^{-2}$ and so can be important in disk physics. 
By fitting both ROSAT and ASCA spectra, an
X-ray luminosity of $L_x$ $\approx 2\times 10^{30}$~ergs~s$^{-1}$ 
was derived for TW~Hya by \citet{kastner_h99}, with prominent 
line emission at energies $\sim$ 1 keV. With these X-ray parameters, 
Glassgold et al. (2004) calculated the ionization rate at the top of the 
disk atmosphere at a radial distance of 1 AU to be 
$\sim6\times 10^{-9}$~s$^{-1}$.  The X-ray luminosity of TW Hya 
is known to be highly variable (Kastner et al. 1999, 2002) due to 
flaring events, but the X-ray ionization important for heating should be 
a time-averaged effect. 
The model calculations concerning X-ray ionization and its effects 
on the gas temperature are subject to uncertainty, since the hard 
X-ray flux (most important for penetrating the disk) assumed in the
Glassgold et al. treatment is a simplified representation of the 
real X-ray emission, which when examined at low resolution is actually 
broadly distributed from a few tenths to several keV 
(Kastner et al. 1999; Stelzer \& Schmitt 2004).

Motivated by the X-ray observations, we focus on X-ray effects
on the disk surface to describe the additional surface heating. 
In order to model the effects of X-rays in heating the gas in 
the upper atmosphere of the TW~Hya disk,
we follow the Glassgold et al. (2001, 2004) scheme
to estimate the gas temperature for the region of the disk 
probed by CO rotational lines. 
The X-ray heating at the disk surface is expressed as
\begin{equation}
\Gamma _{X}=4.8\times 10^{-11}erg \zeta _{X}\left( \frac{\Delta
\epsilon_{h}}{30eV}\right) n_{H}
\end{equation}
in terms of a mean heating energy per ionization, $\Delta\epsilon_h$
and the X-ray ionization rate 
$\zeta _{X}=6\times 10^{-9}s^{-1}\left( \frac{AU}{r}\right) ^{2}$,
which is adopted from Glassgold et al. 2004
for TW~Hya where r is the cylindrical radius from the disk axis.

The dust-gas cooling rate is taken to be
\begin{equation}
\Lambda _{dg}=2\times 10^{33}erg cm^{3} s^{-1}\left\lceil
\frac{\frac{\rho _{d}}{\rho _{g}}}{0.01}\frac{500 {\rm \AA}}{a}\frac{\alpha
}{0.5}\right\rceil T^{0.5}\left(T-T_{d}\right) n_{H}^{2}
\end{equation}
where $\alpha$ is the accommodation coefficient (from equation 3 of 
Glassgold et al. 2004), T is the gas-kinetic
temperature, and T$_d$ is the mean dust temperature. When $T>T_d$ the dust
cools the gas. The net volumetric gain rate of gas energy due to
collisions with dust is then proportional to the local dust-to-gas
ratio $\frac{\rho _{d}}{\rho _{g}}$ and inversely proportional to the
mean dust radius $a$. With the heating of the gas by X-rays 
balanced by the dust-gas cooling rate, 
we can calculate the difference of gas and dust
temperature at each locale within the disk. Figure 3 shows the
gas$-$dust temperature differences overlaid 
with the dust temperature contours from the 
TW~Hya model (Calvet et al. 2002). By incorporating these temperature
difference into the disk model,
we can fit the CO lines using the $\chi^2$ analysis 
in the ($uv$) plane to estimate the goodness of fit for the various 
disk parameters,
including the outer radius R$_{out}$ and the inclination angle $i$,
$(vsini)_{100AU}$. 
A similar $\chi^2$ analysis to determine best fit disk parameters 
was carried out on the disk
of DM Tau by Guilloteau and Dutrey (1998). The $\chi^2$ analysis
in the visibility plane is
essential to avoid the non-linear effects of deconvolution
in the imaging process. 

We adopt the disk physical density and temperature structure 
as derived by Calvet et al. (2002), where we
take the disk temperature as the dust temperature, and we use the 
difference between the gas and dust temperature 
derived by balancing the relevant heating and cooling processes.
We produce a grid of models with a range of various disk
parameters and depletion schemes to simulate the disk as imaged by
a telescope with the resolution constrained only by the grid
sampling (typically of order 5-10 AU in the outer disk,
or $0\farcs1-0\farcs2$). We use a 2D Monte Carlo model
(\citealp{hogerheijde_v00}) to calculate the radiative transfer and
molecular excitation, and we produce simulated observations of the 
model disks with the MIRIAD software package using the UVMODEL routine
to select synthetic visibility observations at the observed $(u,v)$
spacings. Then the $\chi^2$ distance is calculated between the
simulated visibility and the CO J=3--2 and J=2--1 data with 
sufficient signal-to-noise. The best fit
parameters as derived by minimizing the $\chi^2$ distance
are shown in Table 2 (details to be presented in a future paper).
The simulated spectra of the CO lines including X-ray heating are shown 
in Figure 2 with red solid lines. Including X-ray heating improves
the agreement of the model with the data.  Note that the CO J=6--5 
model prediction is based only on the lower transition lines. 

The value $(\frac{\rho _{d}}{\rho _{g}})/a$ is constrained by the
$\chi^2$ fit to be around
$4\times10^{-4}$ with $a$ in $\micron$. If the local dust-to-gas ratio
is around 0.01, then the mean dust radius is around 25$\micron$, which
seems large for the surface of the disk. If dust settles in the mid-plane,
then the local dust-to-gas ratio might be lower, and if it were as low 
as as 0.001, then the mean dust radius would be around micron size or 
smaller, which seems more plausible.

Even though the disk model with X-ray heating can effectively 
match the various CO observations of TW~Hya,
especially the line intensity at $\sim$2$''$ resolution, 
the predicted width of CO J=6--5 line is nearly 50\%  larger 
than observed. 
In the model, the CO J=6--5 emission originates mostly from the upper layers 
of the inner 60 AU of the disk, consistent with the double-peaked 
appearance in the model spectra, as the density of the outer disk is 
so low that CO J=6--5 becomes sub-thermally excited. 
The narrow observed linewidth of CO J=6--5 suggests that most of the
emission is actually coming from the outer part of the disk (less
contribution from the velocity projection from the regions with high Keplerian
rotation). There are at least two scenarios that could account for this:
(1) There is not much CO in the inner disk. Although it is possible to 
make the model CO J=6--5 spectra narrower by blanking out the CO emission 
from the inner 20 AU, it would then be hard to fit both the CO J=3--2 and 2--1
lines simultaneously, and also begs the question as to the origin of the
CO M-band $v$=1--0 emission (\citealp{rettig_h04}). 
(2) The gas temperature of CO in the inner 20 AU is actually lower than 
predicted by the models, possibly due to either additional cooling or 
perhaps shadowing induced by a puffed up inner rim of the disk, 
so that the emission from the higher excitation J=6--5 line is suppressed 
but with little effect on the lower excitation J=3--2 and 2--1 lines.
Clearly the current model is still too simplified to fit all the CO
lines perfectly. Further analysis of the thermal structure in the
inner disk will hopefully help to resolve this puzzle.

\section{Summary}

We have presented the first imaging of the CO J=6--5 emission from the disk around 
a classical T Tauri star, TW~Hya. Using a 2D Monte Carlo simulation with 
a physical description of the disk and an idealized X-ray heating model, 
we are able to simultaneously reproduce the line intensities of 
CO J=6--5, J=3--2 and J=2--1 molecular line emission.  
Some puzzles remain, however, in particular the observed CO J=6-5 
line width that appears to be narrower than predicted by this model.

Circumstellar disks evolve from gas-rich structures with 1\%
of the total mass in dust to disks that appear to be
completely devoid of gas and dominated by emission from the 
dust particles generated by planetesimal collisions. 
Theoretical studies indicate short gas dispersal timescales
(\citealp{hollenbach_j00}). Studies of the gas component of the disk
are needed to obtain detailed information on the mechanisms of disk dispersal.
SMA observations of multiple CO transitions toward TW Hya show that
additional gas heating is needed at the surface of the disk 
to explain the measured line intensities and ratios. 
Detailed modeling with realistic X-ray ionization rates 
derived from the observed X-ray spectrum of 
TW Hya, as well as other surface heating processes (especially 
UV heating) will be essential to understand the nature 
of the gas in the disk of TW Hya, and hence the gas 
evolution in protoplanetary disks in general.

We wish to thank the referee J.H. Kastner for very useful comments.  
C.Q. are grateful for several discussions with A. Glassgold and
S. Shang. M.R.H's research is supported by a VIDI grant from the Netherlands
Organization for Scientific Research (NWO). We thank P. D'Alessio for 
providing the disk model of TW~Hya. 

\clearpage

\bibliographystyle{apj}

\begin{thebibliography}{20}
\expandafter\ifx\csname natexlab\endcsname\relax\def\natexlab#1{#1}\fi

\bibitem[{{Beckwith}(1999)}]{beckwith99}
{Beckwith}, S. V.~W. 1999, in NATO ASIC Proc. 540: The Origin of Stars and
  Planetary Systems, 579

\bibitem[Calvet et al.(2002)]{2002ApJ...568.1008C} Calvet, N., D'Alessio, 
P., Hartmann, L., Wilner, D., Walsh, A., \& Sitko, M.\ 2002, \apj, 568, 
1008 
 

\bibitem[{{Chiang} \& {Goldreich}(1997)}]{cg97}
{Chiang}, E.~I. \& {Goldreich}, P. 1997, \apj, 490, 368

\bibitem[{{Dartois} {et~al.}(2003){Dartois}, {Dutrey}, \&
  {Guilloteau}}]{dartois_d03}
{Dartois}, E., {Dutrey}, A., \& {Guilloteau}, S. 2003, \aap, 399, 773

\bibitem[{{Duvert} {et~al.}(2000){Duvert}, {Guilloteau}, {M{\'e}nard}, {Simon},
  \& {Dutrey}}]{duvert_g00}
{Duvert}, G., {Guilloteau}, S., {M{\'e}nard}, F., {Simon}, M., \& {Dutrey}, A.
  2000, \aap, 355, 165

\bibitem[{{Feigelson} \& {Montmerle}(1999)}]{feigelson_m99}
{Feigelson}, E.~D. \& {Montmerle}, T. 1999, \araa, 37, 363

\bibitem[{{Glassgold} {et~al.}(2004){Glassgold}, {Najita}, \&
  {Igea}}]{glassgold_n04}
{Glassgold}, A.~E., {Najita}, J., \& {Igea}, J. 2004, \apj, 615, 972

\bibitem[{{Glassgold} \& {Najita}(2001)}]{glassgold_n01}
{Glassgold}, A.~E. \& {Najita}, J.~R. 2001, in ASP Conf. Ser. 244: Young Stars
  Near Earth: Progress and Prospects, 251

\bibitem[{{Guilloteau} \& {Dutrey}(1998)}]{guilloteau_d98}
{Guilloteau}, S. \& {Dutrey}, A. 1998, \aap, 339, 467

\bibitem[{{Ho} {et~al.}(2004){Ho}, {Moran}, \& {Lo}}]{ho_m04}
{Ho}, P.~T.~P., {Moran}, J.~M., \& {Lo}, K.~Y. 2004, \apjl, 616, L1

\bibitem[{{Hogerheijde} \& {van der Tak}(2000)}]{hogerheijde_v00}
{Hogerheijde}, M.~R. \& {van der Tak}, F. F.~S. 2000, \aap, 362, 697

\bibitem[{{Hollenbach} {et~al.}(2000){Hollenbach}, {Yorke}, \&
  {Johnstone}}]{hollenbach_j00}
{Hollenbach}, D.~J., {Yorke}, H.~W., \& {Johnstone}, D. 2000, Protostars and
  Planets IV, 401

\bibitem[{{Igea} \& {Glassgold}(1999)}]{igea_g99}
{Igea}, J. \& {Glassgold}, A.~E. 1999, \apj, 518, 848

\bibitem[{{Jonkheid} {et~al.}(2004){Jonkheid}, {Faas}, {van Zadelhoff}, \& {van
  Dishoeck}}]{Jonkheid_f04}
{Jonkheid}, B., {Faas}, F.~G.~A., {van Zadelhoff}, G.-J., \& {van Dishoeck},
  E.~F. 2004, \aap, 428, 511

\bibitem[{{Kastner} {et~al.}(1999){Kastner}, {Huenemoerder}, {Schulz}, \&
  {Weintraub}}]{kastner_h99}
{Kastner}, J.~H., {Huenemoerder}, D.~P., {Schulz}, N.~S., \& {Weintraub}, D.~A.
  1999, \apj, 525, 837

\bibitem[Kastner et al.(2002)]{2002ApJ...567..434K} Kastner, J.~H., 
Huenemoerder, D.~P., Schulz, N.~S., Canizares, C.~R., \& Weintraub, D.~A.\ 
2002, \apj, 567, 434 
 

\bibitem[{{Koerner} \& {Sargent}(1995)}]{koerner_s95}
{Koerner}, D.~W. \& {Sargent}, A.~I. 1995, \aj, 109, 2138

\bibitem[{{Qi} {et~al.}(2004){Qi}, {Ho}, {Wilner}, {Takakuwa}, {Hirano},
  {Ohashi}, {Bourke}, {Zhang}, {Blake}, {Hogerheijde}, {Saito}, {Choi}, \&
  {Yang}}]{qi_h04}
{Qi}, C., {Ho}, P.~T.~P., {Wilner}, D.~J., {Takakuwa}, S., {Hirano}, N.,
  {Ohashi}, N., {Bourke}, T.~L., {Zhang}, Q., {Blake}, G.~A., {Hogerheijde},
  M., {Saito}, M., {Choi}, M., \& {Yang}, J. 2004, \apjl, 616, L11

\bibitem[{{Qi} {et~al.}(2003){Qi}, {Kessler}, {Koerner}, {Sargent}, \&
  {Blake}}]{qi_k03}
{Qi}, C., {Kessler}, J.~E., {Koerner}, D.~W., {Sargent}, A.~I., \& {Blake},
  G.~A. 2003, \apj, 597, 986

\bibitem[{{Rettig} {et~al.}(2004){Rettig}, {Haywood}, {Simon}, {Brittain}, \&
  {Gibb}}]{rettig_h04}
{Rettig}, T.~W., {Haywood}, J., {Simon}, T., {Brittain}, S.~D., \& {Gibb}, E.
  2004, \apjl, 616, L163

\bibitem[Stelzer \& Schmitt(2004)]{2004A&A...418..687S} Stelzer, B., \& 
Schmitt, J.~H.~M.~M.\ 2004, \aap, 418, 687 

\bibitem[van Zadelhoff et al.(2001)]{2001A&A...377..566V} van Zadelhoff, 
G.-J., van Dishoeck, E.~F., Thi, W.-F., \& Blake, G.~A.\ 2001, \aap, 377, 
566 
 
\bibitem[{{Zuckerman} {et~al.}(1995){Zuckerman}, {Forveille}, \&
  {Kastner}}]{zuckerman_f95}
{Zuckerman}, B., {Forveille}, T., \& {Kastner}, J.~H. 1995, \nat, 373, 494

\end{thebibliography}

\clearpage

\begin{figure}
\centering
\includegraphics[width=5in, angle=-90]{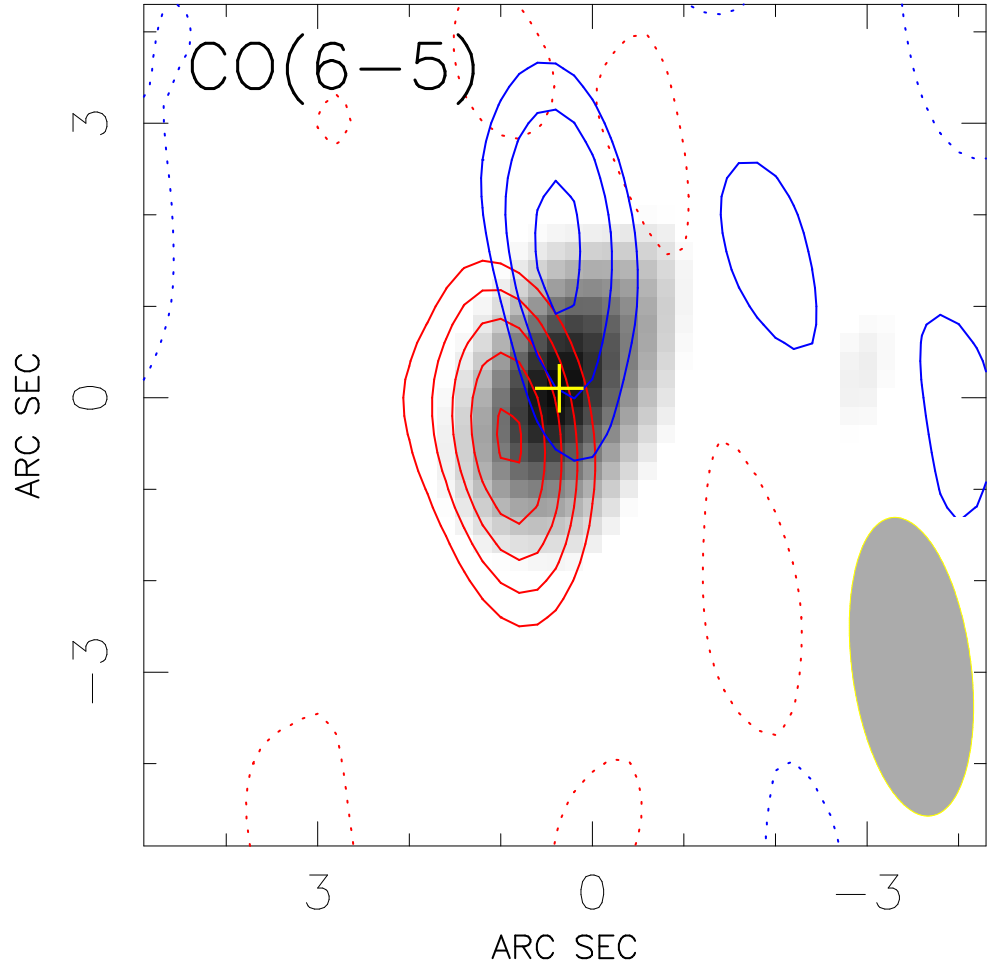}
\caption{CO J=6--5 velocity channel maps ({\it red: 2.9 km s$^{-1}$, blue:
2.5 km s$^{-1}$}) from TW~Hya, overlaid on the 690 GHz dust continuum
map ({\it gray scale}). The cross indicates the position of the
continuum peak.
 \label{fig:image}}
\end{figure}

\clearpage

\begin{figure}
\centering
\includegraphics[width=3.5in]{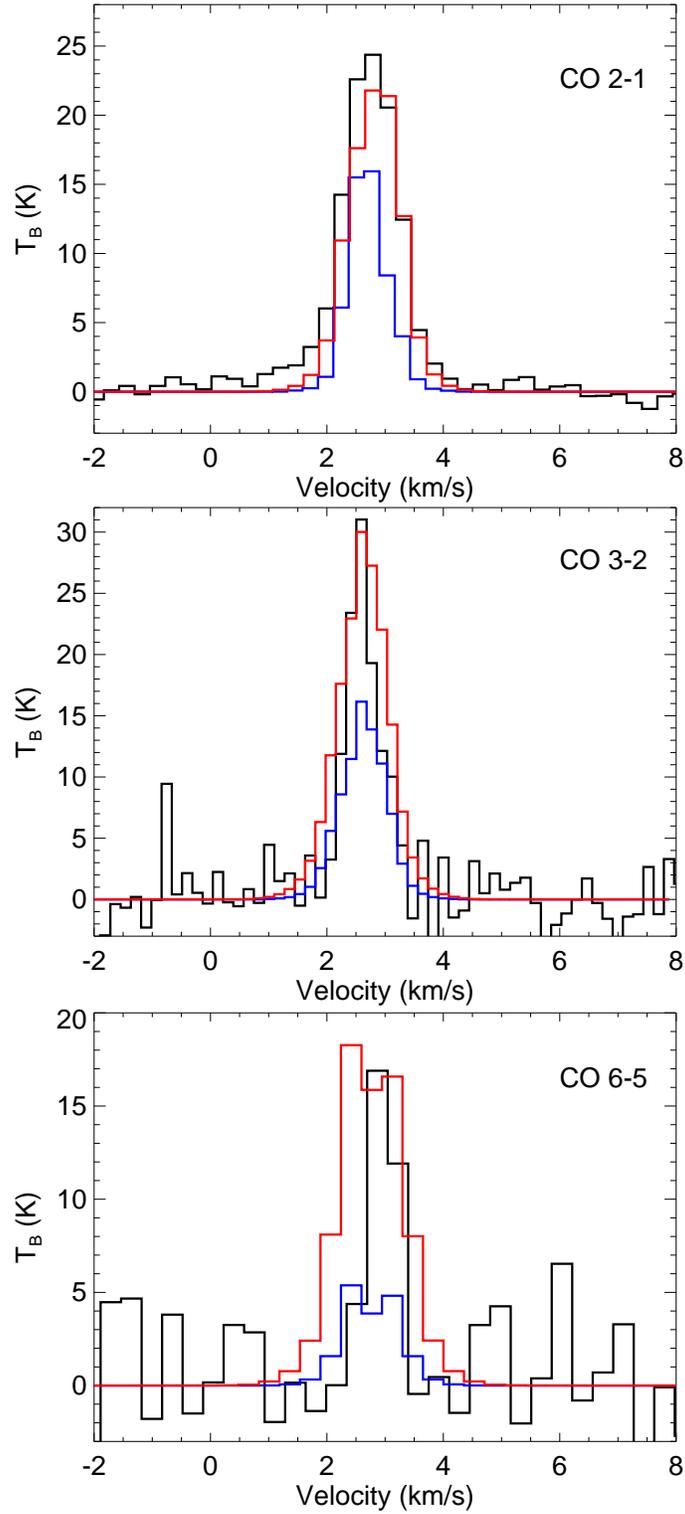}
\caption{ The CO J=2--1, 3--2 and 6--5 spectra at the continuum
(stellar) position. The spectra in black are the SMA data, and the red
and blue spectra are the simulated models with and without X-ray heating.
 \label{fig:spectra}}
\end{figure}

\clearpage

\begin{figure}
\centering
\includegraphics[width=5.0in,angle=90]{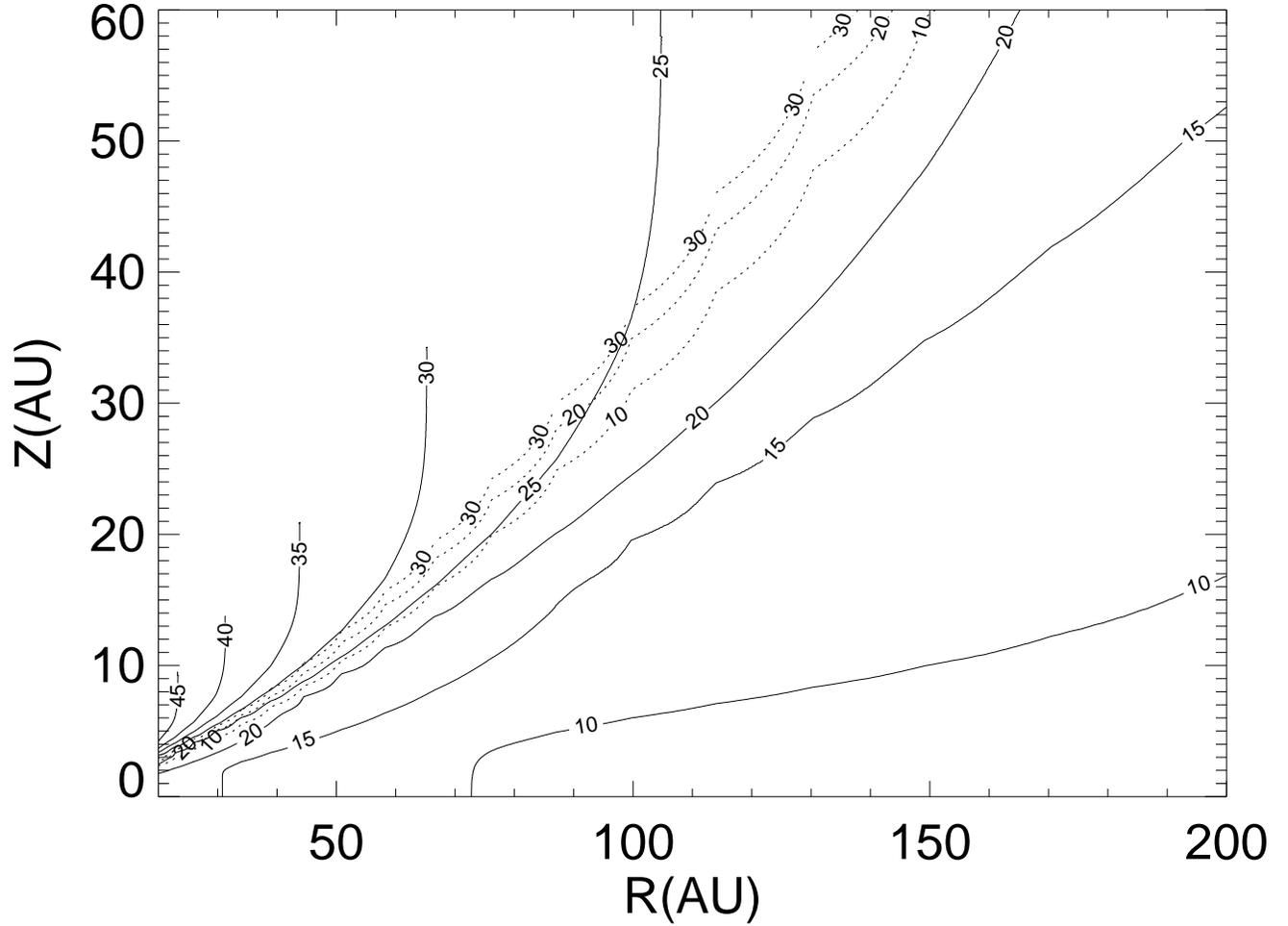}
\caption{The dust temperature contours (in solid lines) from the TW
Hya model (Calvet et al. 2002) and the gas$-$dust temperature
differences (in dotted lines) due to the X-ray gas heating on
the disk surface.}
\end{figure}

\clearpage

\begin{deluxetable}{lcccc}
\tablewidth{0pt}
\tablecaption{Observational Parameters for SMA TW~Hya \label{tab:obs}}
\tablehead{
\colhead{}&\colhead{CO 3--2}&\colhead{CO 2--1}&\colhead{CO 6--5}}
\startdata
Rest Frequency:& 345.796 GHz & 230.538 GHz & 691.473 GHz \\
Synthesized beam: & $2\farcs7 \times 1\farcs6$ PA 18.7$^{\circ}$
                  & $2\farcs7 \times 1\farcs7$ PA 9.9$^{\circ}$
                  & $3\farcs3 \times 1\farcs3$ PA 7.5$^{\circ}$ \\
R.M.S$\tablenotemark{a}$ (continuum): &   35 mJy/beam  &  1.8 mJy/beam
& 110 mJy/beam \\
Dust flux:    & 1.62 $\pm$ 0.05 Jy & 0.54 $\pm$ 0.03 Jy & 4.62 $\pm$
0.54 Jy \\
Channel spacing: & 0.18 km s$^{-1}$ & 0.26 km s$^{-1}$ & 0.35 km
s$^{-1}$ \\
R.M.S.$\tablenotemark{a}$ (line):   & 1.0 Jy/beam & 0.11 Jy/beam & 5.3
Jy/beam \\
Peak intensity & 31.0 K & 24.4 K & 16.9 K \\
\enddata
\tablenotetext{a} { SNR limited by the dynamic range. }
\end{deluxetable}

\clearpage

\begin{deluxetable}{lccc}
\tablewidth{0pt}
\tablecaption{Model Parameters Used in Simulating TW~Hya emission \label{tab:model}}
\tablehead{
\colhead{} &\colhead{Parameters}}
\startdata
Physical Structure & Irradiated accretion disk (Calvet et al. 2002)\\
Stellar Mass & 0.77 M$_{\odot}$\\
Disk Size & R$_{in}$ 4 AU, R$_{out,edge}$ 172 AU\\
Disk PA   & $-$27.4$^{\circ}$ \\
Inclination Angle & 6$^{\circ}$ \\
Turbulent Velocity & 0.12 km s$^{-1}$\\
CO Depletion Factor & Vc=10$^{-1.2}$, (Vj=10$^{-2.2}$ for T$\le$22 K)\\
\enddata
\end{deluxetable}

\end{document}